\documentclass[11pt]{article}

\usepackage[margin=1in]{geometry}
\usepackage{amsmath,amssymb,amsfonts}
\usepackage{graphicx}
\usepackage{booktabs}
\usepackage{natbib}
\usepackage{hyperref}
\usepackage{subcaption}
\usepackage[ruled,noend]{algorithm2e}
\usepackage{tikz}
\usepackage{xcolor}
\usepackage{pgfplots}
\pgfplotsset{compat=1.16}
\usepackage{multirow}
\usepackage{enumitem}

\usetikzlibrary{arrows.meta, positioning, fit, backgrounds, calc, shapes.geometric}

\title{Real-Time Parallel Counterfactual Regret Minimization}
\author{Boning Li\thanks{IIIS, Tsinghua University. Email: li-bn22@mails.tsinghua.edu.cn} 
\and Longbo Huang\thanks{IIIS, Tsinghua University. Email: longbohuang@tsinghua.edu.cn. Corresponding author.}}

\date{\today}

\begin{document}
\maketitle

\begin{abstract}
Counterfactual Regret Minimization (CFR) is the dominant algorithmic family for solving large imperfect-information games, underpinning breakthroughs such as Libratus and Pluribus in No-Limit Texas Hold'em poker. In real-time game-playing systems, the solver must compute a near-equilibrium strategy within a strict time budget of only a few seconds per decision, and the number of CFR iterations completed in this window directly determines play strength. We present \textbf{Parallel CFR}, the first parallelization framework for real-time depth-limited CFR solving that seamlessly integrates pruning, abstraction, and advanced CFR variants. We decompose each CFR iteration into a pipeline of seven stages and identify two orthogonal dimensions of parallelism: \emph{by information set} and \emph{by tree node}. Leaf node evaluation is offloaded to GPUs via batched neural network inference, creating a heterogeneous CPU--GPU pipeline. Experiments on Heads-Up No-Limit Texas Hold'em demonstrate that Parallel CFR achieves $3.3$--$3.4\times$ speedup over the single-threaded baseline on postflop streets, with per-iteration time of ${\sim}47$--$54$~ms on a depth-limited game tree with over $1$ billion histories. All experiments run on a single desktop-class device (NVIDIA DGX Spark), enabling hundreds of CFR iterations within a typical real-time decision budget without requiring datacenter-scale infrastructure.


\end{abstract}

\section{Introduction}
\label{sec:intro}

Imperfect-Information Extensive-Form Games (IIEFGs), where players must make decisions without full knowledge of the game state, model fundamental problems in strategic reasoning, from poker \citep{kuhn1950simplified} to security, auctions, and negotiation.
Solving such games requires computing a \emph{Nash equilibrium}: a strategy profile from which no player can unilaterally deviate to improve their expected payoff.
The Counterfactual Regret Minimization (CFR) family of algorithms \citep{zinkevich2007regret} has been the dominant approach to computing approximate Nash equilibria in large IIEFGs, enabling superhuman performance in poker via systems such as DeepStack \citep{moravvcik2017deepstack}, Libratus \citep{brown2018superhuman} and Pluribus \citep{brown2019superhuman}.

Parallelization has been a key driver of progress in artificial intelligence, enabling the training of large neural networks across thousands of GPUs. In contrast, parallelization of game-solving algorithms has received remarkably little attention. Existing work has explored parallelizing tabular CFR \citep{johanson2011accelerating,johanson2012efficient}, but these techniques cannot be directly applied to the modern real-time game-solving stack, which combines depth-limited solving \citep{moravvcik2017deepstack}, pruning \citep{li2025evpa}, abstraction \citep{li2026effective}, and advanced CFR variants \citep{brown2019solving,farina2021faster} into a heterogeneous computational pipeline that resists simple parallelization. Yet each component contains substantial internal parallelism that has not been systematically exploited.

This parallelization gap has direct practical consequences.
In real-time game-playing systems such as Libratus \citep{brown2018superhuman} and Pluribus \citep{brown2019superhuman}, the solver must compute a near-equilibrium strategy at each decision point within a strict time budget of only a few seconds.
The quality of the resulting strategy is directly determined by how many CFR iterations the solver can complete within this window.
A faster per-iteration pipeline therefore translates directly into stronger play, making parallelization not merely an engineering convenience but a fundamental driver of AI performance.
Moreover, previous superhuman poker AI systems often relied on substantial computational resources for real-time solving (e.g., Libratus \citep{brown2018superhuman} used a supercomputer during real-time play).
An efficient parallel pipeline that exploits modern heterogeneous hardware can bring real-time solving to compact, consumer-grade devices, dramatically lowering the deployment barrier for game-solving AI.

In this paper, we present \textbf{Parallel CFR}, a comprehensive parallelization framework for practical CFR-based game solving.
Experiments on Heads-Up No-Limit Texas Hold'em demonstrate that Parallel CFR achieves $3.3$--$3.4\times$ speedup on postflop streets over the single-threaded baseline using only 5 CPU threads and 1 GPU on a single NVIDIA DGX Spark (a desktop-class device), with per-iteration pipeline time of ${\sim}47$--$54$~ms on a depth-limited game tree with over $1$ billion histories.
A convergence comparison against PokerRL~\citep{steinberger2019pokerrl}, an efficient open-source poker solver, further confirms that Parallel CFR reaches $\sim7\times$ lower exploitability at the same wall-time budget.
Our key contributions are:

\begin{itemize}
    \item We propose Parallel CFR, the \textbf{first parallelization framework for real-time depth-limited CFR solving}.
    We decompose a single CFR iteration into a seven-stage pipeline and identify two orthogonal dimensions of parallelism: \emph{by information set} and \emph{by tree node}.
    Each stage is independently parallelizable along one or both dimensions via multi-threaded CPU execution.

    \item We design a \textbf{heterogeneous CPU--GPU pipeline} that natively integrates all essential components of practical game solving, including depth-limited solving, pruning, information abstraction, and advanced CFR variants, into a unified parallel architecture.
    Leaf node evaluations are batched into a single GPU forward pass, and pruning acts as a multiplicative accelerator on all pipeline stages.

    \item Parallel CFR achieves the \textbf{fastest known CFR solving speed}, enabling real-time solving on a single desktop-class device. Parallel CFR completes each CFR iteration in ${\sim}47$--$54$~ms on the depth-limited game tree with over $1$ billion histories, allowing hundreds of iterations within a typical real-time decision budget of a few seconds. A direct comparison against an efficient poker solver further shows that Parallel CFR reaches $\sim7\times$ lower exploitability at equal wall-clock time.
\end{itemize}

\section{Related Work}
\label{sec:related}

\paragraph{CFR algorithms.}
CFR \citep{zinkevich2007regret} introduced regret-based self-play for imperfect-information games.
Subsequent variants improved convergence: CFR$^+$ \citep{tammelin2014solving} floors negative regrets, DCFR \citep{brown2019solving} applies asymmetric discounting, and Predictive CFR$^+$ \citep{farina2021faster} leverages predictive regret matching. All of these are iterative algorithms operating on a fixed game tree, making them natural targets for parallelization.

\paragraph{Depth-limited solving.} DeepStack \citep{moravvcik2017deepstack} introduced neural network-based depth-limited solving for HUNL. Modicum \citep{brown2018depth} proposed a state-based depth-limited method. ReBeL \citep{brown2020combining} extended this to a general RL+Search framework and formalized the approach and established theoretical guarantees. Recent work has focused on accelerating the training of depth-limited solving \citep{li2026turborebel}, and our implementation uses the same method to train the leaf estimation network. Notably, EVPA \citep{li2025evpa} has implemented online CFR pruning and abstraction in depth-limited solving, reducing the online solving overhead by up to two orders of magnitude. These systems implicitly use GPU for neural network inference but do not systematically parallelize the CFR tree operations themselves.

\paragraph{Abstraction.} Abstract techniques combine multiple sets of information in a game tree into a bucket and take the same action, thereby significantly reducing the size of the game tree. Abstraction can be divided into information abstraction, action abstraction, and imperfect-recall abstraction. Information abstraction groups similar information sets into \emph{buckets} \citep{gilpin2006competitive}, which can be divided into lossless abstraction and lossy abstraction. Lossless abstraction requires judgment based on game rules \citep{gilpin2007lossless}, while lossy abstraction can further reduce solving costs. If we use lossy abstraction, we will use the state-of-the-art WEVA algorithm \citep{li2026effective}. Action abstraction selects a small number of representative actions from a large number of legal actions to construct a game tree \citep{hawkin2011automated,li2024rl}, which is a necessary abstraction in games such as HUNL. We use fixed action abstractions to ensure fair comparisons. The imperfect-recall abstraction allows players to forget some of the game's common knowledge, reducing the size of the super large game tree \citep{waugh2009practical,ganzfried2014potential}. It is an offline method for calculating full game blueprint strategy \citep{brown2015hierarchical}, independent of our online depth-limited solving method.


\paragraph{Pruning.} Pruning has proven to be one of the most effective techniques for accelerating CFR convergence, often achieving speedups of up to two orders of magnitude compared to unpruned variants. 
\emph{Regret-based pruning} \citep{brown2015regret} skips subtrees that are unreachable under the current strategy profile. \emph{Best-response pruning} \citep{brown2017reduced} temporarily eliminates actions that are dominated with respect to an approximate best response.
\emph{Dynamic thresholding pruning} \citep{brown2017dynamic} further extend these ideas by pruning low-probability actions on the fly, adapting to the current regret state. Most recently, \emph{EVPA} \citep{li2025evpa} proposed a pruning method that can be used for depth-limited solving, and our pruning implementation adopts this approach.


\paragraph{Parallelizing game solving.} Prior parallelization efforts in CFR fall into several distinct categories, none of which parallelize across different information sets within a single iteration.
\emph{Subtree decomposition} \citep{burch2014solving} splits a large game into smaller subgames that can be solved independently, providing coarse-grained parallelism across machines but no parallelism within each subgame solve. 
\emph{Node-level parallel traversal} for tabular CFR was explored by \citet{johanson2011accelerating} in the context of best-response computation, parallelizing the tree walk over disjoint subtrees rooted at a single depth. \emph{MCCFR sampling parallelism} \citep{lanctot2009monte,johanson2012efficient} could run independent sampled trajectories in parallel across workers, but each trajectory still traverses the tree serially and high variance from sampling limits practical speedup. More recently, \citet{kim2026parallelizingcounterfactualregretminimization} proposed reformulating sequence-form CFR as linear algebra operations to leverage GPU acceleration.
Although their method has some acceleration compared to the most primitive CFR implementation in fully enumerable games, it does not support basic components in real-time game solving pipelines such as depth-limited solving and  pruning, and its parallel implementation on GPU is much slower than the most advanced CFR implementation methods on single threaded CPU \citep{steinberger2019pokerrl} (one of our baseline).

Crucially, all prior work shares two fundamental limitations that our Parallel CFR addresses:
\textbf{(1)} None of these methods are designed for \emph{depth-limited subgames} nor can they be seamlessly combined with \emph{pruning} \citep{li2025evpa}; and
\textbf{(2)} Their parallelization granularity is tied to \emph{tree depths or public nodes}, rather than enabling independent parallelism \emph{across information sets}, which is the natural unit of computation in CFR.
In contrast, Parallel CFR introduces a seven-stage pipeline that parallelizes \emph{by information set} and \emph{by node}, integrates pruning and abstraction as a multiplicative accelerator, and operates natively on depth-limited game trees with neural leaf evaluation.

\section{Notation}
\label{sec:notation}

An \emph{Imperfect Information Extensive-Form Game} (IIEFG) is defined by a tuple $\langle \mathcal{N}, \mathcal{H}, \mathcal{Z}, \mathcal{A}, u, \mathcal{I} \rangle$, where $\mathcal{N} = \{1, \ldots, N\}$ is the set of players, $\mathcal{H}$ is the set of non-terminal histories (game states), $\mathcal{Z}$ is the set of terminal histories, $\mathcal{A}(h)$ is the set of actions available at history $h$, $u_i: \mathcal{Z} \to \mathbb{R}$ is the utility function for player $i$, and $\mathcal{I}_i$ is the partition of player $i$'s histories into \emph{information sets}.
An information set $I \in \mathcal{I}_i$ groups all histories that player $i$ cannot distinguish.

A \emph{behavioral strategy} $\sigma_i$ assigns a probability distribution over actions at each information set: $\sigma_i(I, a) = \Pr[\text{player } i \text{ plays } a \text{ at } I]$.
A \emph{strategy profile} $\sigma = (\sigma_1, \ldots, \sigma_N)$ specifies a strategy for each player.

The \emph{reach probability} $\pi^\sigma(h)$ is the probability of reaching history $h$ under strategy profile $\sigma$.
It decomposes as $\pi^\sigma(h) = \pi^\sigma_i(h) \cdot \pi^\sigma_{-i}(h) \cdot \pi^\sigma_c(h)$, where $\pi^\sigma_i(h)$ is the product of player $i$'s action probabilities on the path to $h$, $\pi^\sigma_{-i}(h)$ is the product of all opponents' action probabilities, and $\pi^\sigma_c(h)$ is the product of chance probabilities.

The \emph{counterfactual value} (CFV) of an information set $I$ for player $i$ under strategy profile $\sigma$ is:
\begin{equation}
    v_i(\sigma, I) = \sum_{h \in I} \pi^\sigma_{-i}(h) \sum_{z \in \mathcal{Z}(h)} \pi^\sigma(h, z) \cdot u_i(z),
    \label{eq:cfv}
\end{equation}
where $\mathcal{Z}(h)$ is the set of terminal histories reachable from $h$, and $\pi^\sigma(h, z)$ is the probability of transitioning from $h$ to $z$ under $\sigma$.

The \emph{instantaneous counterfactual regret} for action $a$ at information set $I$ is:
\begin{equation}
    r^t_i(I, a) = v_i(\sigma^t_{I \to a}, I) - v_i(\sigma^t, I),
\end{equation}
where $\sigma^t_{I \to a}$ is the strategy that plays action $a$ with probability 1 at $I$ and follows $\sigma^t$ elsewhere.
The \emph{cumulative regret} is $R^T_i(I, a) = \sum_{t=1}^T r^t_i(I, a)$.

CFR \citep{zinkevich2007regret} uses \emph{regret matching} to update the strategy:
\begin{equation}
    \sigma^{t+1}_i(I, a) = \frac{[R^t_i(I, a)]^+}{\sum_{a'} [R^t_i(I, a')]^+},
    \label{eq:rm}
\end{equation}
where $[x]^+ = \max(x, 0)$.
If the denominator is zero, a uniform strategy is used.
The average strategy $\bar{\sigma}^T$ converges to a Nash equilibrium at a rate of $O(1/\sqrt{T})$ in two-player zero-sum games.

In large games, the complete game tree cannot be enumerated.
\emph{Depth-limited solving} \citep{moravvcik2017deepstack} builds the game tree only to a depth limit $D$, creating non-terminal leaf nodes.
At these leaves, a trained neural network $f_\theta$ estimates counterfactual values:
\begin{equation}
    \hat{v}_i(I_\text{leaf}) = f_\theta(\pi^\sigma_1(I_\text{leaf}), \ldots, \pi^\sigma_N(I_\text{leaf}), \text{public state}),
\end{equation}
where the input encodes normalized reach probabilities for all players.

\section{Parallel CFR}
\label{sec:method}

We present Parallel CFR, a framework that decomposes each CFR iteration into a pipeline of seven stages.
The pipeline exploits two dimensions of parallelism, \emph{by information set} and \emph{by tree node}, and combines CPU multi-threading with GPU batch inference.
The entire pipeline is \emph{computationally exact}: it produces the same numerical result as a serial CFR implementation with no approximation error. We first describe the pipeline structure (\S\ref{ssec:overview}), then detail the forward pass (\S\ref{ssec:forward}), parallel middle pass (\S\ref{ssec:middle}), backward pass (\S\ref{ssec:backward}), and integration with pruning and abstraction (\S\ref{ssec:integration}).

\subsection{Pipeline Overview}
\label{ssec:overview}

A CFR iteration involves three types of computation: (1)~propagating reach probabilities through the game tree, (2)~evaluating terminal and leaf nodes, and (3)~backpropagating counterfactual values and updating regrets.
These operations have different data dependencies and parallelism patterns.
We decompose each iteration into seven stages (Table~\ref{tab:pipeline}), so that each stage can use the parallelization strategy best suited to its structure.

\begin{table}[h]
\centering
\small
\caption{Seven-stage CFR iteration pipeline. Each stage is parallelizable along at least one dimension.}
\label{tab:pipeline}
\begin{tabular}{clll}
\toprule
\textbf{Stage} & \textbf{Operation} & \textbf{Parallel Dim.} & \textbf{Data Flow} \\
\midrule
1 & \textsc{ForwardProfile} & Information set & $\sigma \to \pi_i, \pi_{-i}$ \\
2 & \textsc{AggregateProbSum} & Node $\times$ Card & $\pi_i \to P_\text{sum}[n],\, \mathbf{x}_\text{leaf}$ \\
3 & \textsc{Compute $\pi_{-i}(I)$} & Node + Infoset & $P_\text{sum} \to \pi_{-i}(I)$ \\
4 & \textsc{ShowdownEquity} & Node & $\pi_{-i}(I) \to v_\text{sd}$ \\
5 & \textsc{BatchLeafEval} & GPU batch & $\mathbf{x}_\text{leaf} \to \hat{v}_\text{leaf}$ \\
6 & \textsc{BackwardCFV} & Node + Infoset & $\hat{v}_\text{leaf}, v_\text{sd} \to v_i(I)$ \\
7 & \textsc{UpdateRegret} & Information set & $v_i(I) \to R_i(I,a)$ \\
\bottomrule
\end{tabular}
\end{table}

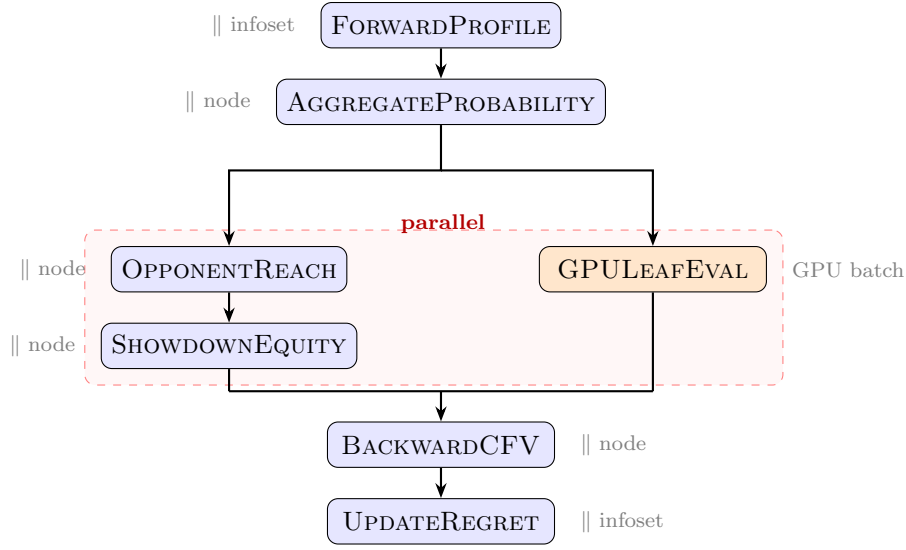
\begin{figure}[t]
\centering
\begin{tikzpicture}[
    stage/.style={draw, rounded corners, minimum width=3.0cm, minimum height=0.6cm, font=\small},
    cpu/.style={stage, fill=blue!10},
    gpu/.style={stage, fill=orange!20},
    arr/.style={-{Stealth[length=2mm]}, thick},
    node distance=0.4cm
]
\node[cpu] (s1) {\textsc{ForwardProfile}};
\node[cpu, below=of s1] (s2) {\textsc{AggregateProbability}};

\coordinate[below=0.6cm of s2] (fork);

\node[cpu, below=1.0cm of fork, xshift=-2.8cm] (s3) {\textsc{OpponentReach}};
\node[cpu, below=of s3] (s4) {\textsc{ShowdownEquity}};

\node[gpu, below=1.0cm of fork, xshift=2.8cm] (s5) {\textsc{GPULeafEval}};

\coordinate[below=0.8cm of $(s4.south)!0.5!(s5.south)$] (join);

\node[cpu, below=0.4cm of join] (s6) {\textsc{BackwardCFV}};
\node[cpu, below=of s6] (s7) {\textsc{UpdateRegret}};

\draw[arr] (s1) -- (s2);

\draw[arr] (s2.south) -- (fork) -| (s3.north);
\draw[arr] (fork) -| (s5.north);

\draw[arr] (s3) -- (s4);

\coordinate (join-l) at (s4.south |- join);
\coordinate (join-r) at (s5.south |- join);
\draw[thick] (s4.south) -- (join-l);
\draw[thick] (s5.south) -- (join-r);
\draw[thick] (join-l) -- (join-r);
\draw[arr] (join) -- (s6);

\draw[arr] (s6) -- (s7);

\node[font=\scriptsize\bfseries, red!70!black] at ($(s3.north east)!0.5!(s5.north west)+(0,0.3)$) {parallel};

\begin{scope}[on background layer]
\node[draw=red!50, dashed, rounded corners, inner sep=6pt,
      fit=(s3)(s4)(s5), fill=red!3] (parbox) {};
\end{scope}

\node[left=0.2cm of s1, font=\scriptsize, gray] {$\parallel$ infoset};
\node[left=0.2cm of s2, font=\scriptsize, gray] {$\parallel$ node};
\node[left=0.2cm of s3, font=\scriptsize, gray] {$\parallel$ node};
\node[left=0.2cm of s4, font=\scriptsize, gray] {$\parallel$ node};
\node[right=0.2cm of s5, font=\scriptsize, gray] {GPU batch};
\node[right=0.2cm of s6, font=\scriptsize, gray] {$\parallel$ node};
\node[right=0.2cm of s7, font=\scriptsize, gray] {$\parallel$ infoset};

\end{tikzpicture}
\caption{Pipeline of a single CFR iteration in Parallel CFR. Blue stages execute on CPU with OpenMP parallelism; the orange stage runs on GPU. Stages 3/4 and Stage 5 execute in parallel (dashed box). Annotations indicate the parallelization dimension.}
\label{fig:pipeline}
\end{figure}

Figure~\ref{fig:pipeline} illustrates the data dependencies.
Stages~1--2 form the \emph{forward pass}, propagating reach probabilities from root to leaves.
After Stage~2, the pipeline \emph{forks}: Stages~3--4 (CPU) and Stage~5 (GPU) execute in parallel since they share no data dependency.
Stages~6--7 form the \emph{backward pass}, beginning after both branches complete.
This decomposition has two advantages: (1)~each stage has a well-defined interface and can be optimized independently; (2)~the structure is modular, since the choice of CFR variant only affects Stage~7 while Stages~1--6 remain identical.

\subsection{Forward Pass (Stages 1--2)}
\label{ssec:forward}

The forward pass propagates reach probabilities from root to leaves and prepares inputs for downstream evaluation.
We operate at the \emph{information-set level} rather than the history level: all hands within an information set share the same strategy, so computation is vectorized over hands within each information set.
However, different hands face different sets of feasible opponent hands due to \emph{card blocking} (two players cannot hold overlapping cards).
This requires additional aggregation structures for efficient blocking correction in later stages.

\paragraph{Stage 1: Forward Profile ($\parallel$ information set).}
Information sets for a given player form disjoint \emph{chains} in the game tree.
Within each chain, reach probabilities propagate sequentially:
\begin{equation}
    \pi_i(I_{\text{child}}) = \pi_i(I_{\text{parent}}) \cdot \sigma_i(I_{\text{parent}}, a),
\end{equation}
where $a$ is the action leading to $I_{\text{child}}$.
Chains are mutually independent, enabling parallel processing across all chains.
The cumulative reach-weighted strategy $\bar{\pi}_i(I)$ is accumulated simultaneously.

\paragraph{Stage 2: Aggregate Probability ($\parallel$ node $\times$ card).}
For each node $n$, we aggregate reach probabilities into three levels (total reach $P_{\text{sum}}$, per-card reach $P_{\text{card}}$, and per-hand reach $P_{\text{hand}}$) that enable $O(1)$ card-blocking corrections in Stage~3 (see Appendix~\ref{app:details} for definitions).
Node-level aggregation is embarrassingly parallel; per-card sums add a second parallel dimension of size $|\mathcal{C}|$ (e.g., 52 in poker).

For depth-limited solving, this stage also constructs the batched input tensor $\mathbf{X} \in \mathbb{R}^{m \times d}$ for all $m$ non-terminal leaf nodes, preparing them for GPU evaluation in Stage~5.
When information abstraction is used, projection matrices bridge the game tree and network representation spaces.

\subsection{Parallel Middle Pass (Stages 3--5)}
\label{ssec:middle}

After the forward pass, the pipeline forks into two independent branches: Stages~3--4 on CPU and Stage~5 on GPU.
Since they share no data dependency, they execute in parallel.
In practice, GPU evaluation time is entirely masked by the longer CPU branch on postflop streets.

\paragraph{Stage 3: Opponent Reach ($\parallel$ node, then $\parallel$ infoset).}
Computing the opponent counterfactual reach $\pi_{-i}(I)$ naively requires summing over all opponent hands at each information set, yielding $O(|\mathcal{H}|^2)$ complexity.
We decompose this into two phases: first, compute per-node multiplicative ratios $\mu[n,h]$ using the three-level aggregates from Stage~2 via an inclusion-exclusion formula (parallel by node); then propagate $\pi_{-i}$ along information set chains using these ratios (parallel by chain).
This avoids the quadratic cost while exposing parallelism in both phases.

\paragraph{Stage 4: Showdown Equity ($\parallel$ node).}
At showdown terminals, each hand's equity must be computed against the opponent's reach-weighted range.
\citet{johanson2011accelerating} observed that hand ranks induce a total order at showdown, enabling a rank-sorted linear scan that reduces per-node complexity from $O(n^2)$ to $O(n)$ where $n$ is the number of hands.
We adopt this technique with our card-blocking corrections.
Showdown nodes are independent, so this stage parallelizes trivially across nodes.
For games lacking the rank-monotonicity property of poker, one may pre-store a payoff matrix and use sparsification methods \citep{farina2022fast} to compute showdown values efficiently.

\paragraph{Stage 5: GPU Leaf Evaluation.}
A single GPU forward pass evaluates all non-terminal leaves simultaneously:
$\hat{\mathbf{V}} = f_\theta(\mathbf{X}) \in \mathbb{R}^{m \times k}$,
where $k$ is the output dimension (hands $\times$ players).
This reduces kernel launch overhead and keeps the GPU well-utilized.
CPU--GPU data transfer occurs exactly twice per iteration (input upload and output download), amortized across all leaf nodes.

\subsection{Backward Pass (Stages 6--7)}
\label{ssec:backward}

The backward pass converts leaf values into regret updates that drive strategy improvement.

\paragraph{Stage 6: Backward CFV ($\parallel$ information set).}
After GPU leaf evaluation returns $\hat{\mathbf{V}}$, counterfactual values at each leaf and chance node are computed by weighting predictions by the opponent's counterfactual reach $\pi_{-i}(I)$ and chance probabilities.
Values then propagate bottom-up within each information set chain:
\begin{equation}
    v_i(\sigma, I) = \sum_{a \in \mathcal{A}(I)} \sigma_i(I, a) \cdot v_i(\sigma, I \cdot a),
    \label{eq:backward}
\end{equation}
where $I \cdot a$ denotes the information set reached after action $a$.
Chains are processed in reverse topological order and are mutually independent.

\paragraph{Stage 7: Update Regret ($\parallel$ information set).}
The instantaneous counterfactual regret for each action is:
\begin{equation}
    r_i(I, a) = v_i(\sigma, I \cdot a) - v_i(\sigma, I),
\end{equation}
where $v_i(\sigma, I \cdot a)$ is the counterfactual value of the child information set after action $a$, as computed in Stage~6.
Updates are purely local to each information set and embarrassingly parallel.
The choice of CFR variant modifies \emph{only} this update rule; Stages~1--6 are identical across all variants.

\subsection{Integration with Pruning and Abstraction}
\label{ssec:integration}
\label{ssec:pruning}
\label{ssec:abstraction}

\paragraph{Online Pruning.}
Our pipeline integrates with online pruning methods that remove dominated actions \emph{before} CFR iterations begin.
We adopt the EVPA framework \citep{li2025evpa}, which uses a neural network ensemble to bound counterfactual values and prune actions whose optimistic value is dominated by a sibling's pessimistic value.
\citet{li2025evpa} proved that this pruning is sound and demonstrated up to two orders of magnitude speedup.
Since pruning is performed once as a preprocessing step, it reduces the effective tree size by a factor $\rho \in [0,1]$ for all subsequent CFR iterations, providing multiplicative speedup to every pipeline stage.
Pruning decisions at each information set are independent, so the pruning pass itself is embarrassingly parallel.
All experiments in this paper enable pruning by default.

\paragraph{Information Abstraction.} Our framework supports both lossless and lossy abstraction. Lossless abstraction introduces \emph{no} precision error and preserves convergence to a Nash equilibrium. All experiments in this paper use lossless abstraction by default. For games with larger hand spaces (e.g., Pot-Limit Omaha with $\binom{52}{4} = 270{,}725$ hands), lossy abstraction becomes necessary. This introduces approximation error inherent to information abstraction itself \citep{johanson2013evaluating}, not to our parallelization. When the neural network operates on a different dimensionality than the game tree, sparse projection matrices bridge the two representation spaces (see Appendix~\ref{app:projection}).



\paragraph{Correctness guarantee.}
Our pipeline produces numerically identical results to a serial CFR implementation: every stage performs the same arithmetic, distributed across threads and devices.
No approximation, sampling, or rounding is introduced by the parallelization.
The sole source of precision loss is lossy information abstraction when optionally enabled, which is a property of the abstraction method itself \citep{johanson2013evaluating}, not of our parallel framework.
Parallel CFR therefore inherits the convergence guarantees of whichever CFR variant is selected in Stage~7.
\section{Experiments}
\label{sec:experiments}
\paragraph{Game.}
HUNL with depth-limited solving to the end of each betting round.
We vary the stack-to-pot ratio $\text{SPR} \in \{4, 16, 64\}$ and the number of raise actions $|\mathcal{A}_\text{raise}| \in \{2, 3\}$, which together control the effective tree size.

\paragraph{Hardware.}
Experiments are conducted on NVIDIA DGX Spark with GB10 Grace Blackwell.
We use OpenMP for CPU parallelism and PyTorch C++ API (LibTorch) for GPU inference.

\paragraph{Baselines.} All configurations use DCFR, pruning, lossless abstraction and depth-limited solving by default.  The default solving setting is a depth-limited game tree with SPR$=64$, $|A_{\textit{raise}}|=3$, which has $2{,}006{,}256$ information sets and $1{,}084{,}381{,}368$ histories.

\paragraph{Metrics.}
Wall-clock time per CFR iteration in milliseconds. 
We report per-street breakdowns: \emph{Preflop}, \emph{Flop}, \emph{Turn}, and \emph{River}.

\paragraph{Overall speedup.}

\begin{table}[h]
\centering
\caption{Per-iteration CFR pipeline time on HUNL.}
\label{tab:overall}
\begin{tabular}{lrrrr}
\toprule
\textbf{Configuration} & \textbf{Preflop} & \textbf{Flop} & \textbf{Turn} & \textbf{River} \\
\midrule
Parallel CFR ($5$ Threads)              & $3.48{\pm}0.22$ & $54.47{\pm}0.40$ & $52.09{\pm}0.25$ & $47.04{\pm}0.25$ \\
Single-thread       & $8.11{\pm}0.02$ & $181.12{\pm}1.54$ & $173.45{\pm}1.30$ & $159.99{\pm}1.69$ \\
\midrule
\textbf{Speedup}               & $2.3\times$ & $3.3\times$ & $3.3\times$ & $3.4\times$ \\
\bottomrule
\end{tabular}
\end{table}

Table~\ref{tab:overall} shows per-iteration pipeline time for the full parallel system and the single-threaded baseline.
Parallel CFR achieves postflop speedups of $3.3\times$ (flop), $3.3\times$ (turn), and $3.4\times$ (river) over the single-threaded baseline.
Preflop is inherently fast (${\sim}3.5$~ms) due to hand isomorphism, yielding a modest $2.3\times$ speedup limited by thread synchronization overhead relative to useful work.
The sub-linear scaling (5 threads, ${\sim}3.3\times$ speedup) is expected: GPU leaf evaluation (Stage~5) is not parallelized across CPU threads, and memory bandwidth becomes a bottleneck as multiple threads compete for shared L3 cache.
Nevertheless, the absolute per-iteration time of $47$--$54$~ms on postflop means that a real-time solver can complete ${\sim}100$ iterations within a 5 second decision budget for a huge subgame with over $1$ billion histories, sufficient for strong approximate equilibrium play with pruning.

\paragraph{Per-stage breakdown.}

Table~\ref{tab:breakdown} breaks down the per-iteration time into the seven pipeline stages across all four streets in a single representative run.
Backward CFV propagation (S6) and opponent counterfactual reach computation (S3) are the two most expensive CPU stages, together accounting for over 50\% of total pipeline time on flop and turn.
GPU leaf evaluation (S5) is the bottleneck for preflop, while CPU stages dominate on flop and turn.
On the river, showdown equity (S4) adds $10.207$~ms and there is no GPU leaf evaluation, making the river purely CPU-bound.

\begin{table}[!ht]
\centering
\caption{Per-stage CFR pipeline time for full configuration.}
\label{tab:breakdown}
\begin{tabular}{lrrrr}
\toprule
\textbf{Stage} & \textbf{Preflop} & \textbf{Flop} & \textbf{Turn} & \textbf{River} \\
\midrule
S1: ForwardProfile & 0.664 & 8.938 & 8.548 & 8.165 \\
S2: AggregateProbability    & 0.914 & 8.276 & 7.649 & 4.999 \\
S3: OpponentReach     & 0.848 & 14.219 & 13.680 & 12.880 \\
S4: Showdown       & 0.000 & 0.000 & 0.000 & 10.207 \\
S5: GPULeafEval        & 2.444 & 11.519 & 11.518 & 0.000 \\
S6: BackwardCFV    & 1.255 & 16.894 & 16.593 & 7.582 \\
S7: UpdateRegret   & 0.447 & 5.333 & 5.067 & 4.815 \\
\midrule
\textbf{Pipeline Total} & 4.196 & 53.660 & 51.536 & 48.648 \\
\bottomrule
\end{tabular}
\end{table}

\begin{figure}[h]
\centering
\begin{tikzpicture}
\begin{axis}[
    width=0.75\columnwidth,
    height=6cm,
    xlabel={Thread count $K$},
    ylabel={Time per iteration (ms)},
    xmin=0.5, xmax=5.5,
    ymin=0, ymax=220,
    xtick={1,2,3,4,5},
    legend pos=north east,
    grid=major,
    legend style={font=\small},
]
\addplot[red, thick, mark=square*] coordinates {(1,8.10) (2,5.17) (3,4.76) (4,4.14) (5,3.54)};
\addlegendentry{Preflop}
\addplot[green!60!black, thick, mark=triangle*] coordinates {(1,181.06) (2,99.64) (3,72.92) (4,62.06) (5,54.64)};
\addlegendentry{Flop}
\addplot[blue, thick, mark=*] coordinates {(1,172.91) (2,95.71) (3,70.25) (4,59.39) (5,52.54)};
\addlegendentry{Turn}
\addplot[orange, thick, mark=diamond*] coordinates {(1,160.42) (2,88.92) (3,64.07) (4,54.77) (5,47.40)};
\addlegendentry{River}
\end{axis}
\end{tikzpicture}
\caption{Per-iteration CFR pipeline time vs.\ thread count $K$.
}
\label{fig:scaling}
\end{figure}
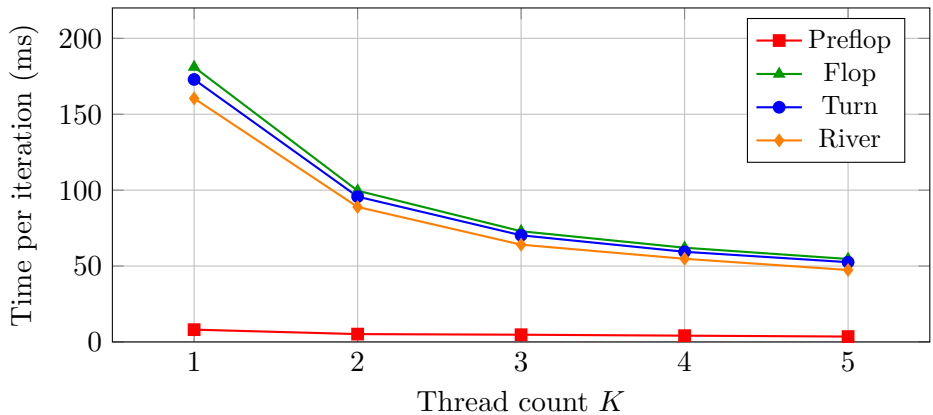

\paragraph{Thread scaling.}
We measure per-iteration time as a function of thread count $K \in \{1, 2, 3, 4, 5\}$ in a same board, pinning threads to 5 big cores within a single L3 cache cluster on DGX Spark.
Figure~\ref{fig:scaling} shows per-street timing.
Threading provides consistent speedup: from $K{=}1$ to $K{=}5$, flop time drops from $181.06$ to $54.64$~ms ($3.31\times$), turn from $172.91$ to $52.54$~ms ($3.29\times$), and river from $160.42$ to $47.40$~ms ($3.38\times$).

\paragraph{Action abstraction and SPR.} Different numbers of raise actions and stack-to-pot ratios (SPR) produce game trees of varying sizes.
We evaluate Parallel CFR performance across $|\mathcal{A}_\text{raise}| \in \{2, 3\}$ and $\text{SPR} \in \{4, 16, 64\}$.
Table~\ref{tab:action_spr} reports per-iteration wall-clock time broken down by street, along with the number of game-tree nodes and maximum information sets per subgame. 
Pipeline time scales approximately linearly with the number of tree nodes, confirming that the parallelization overhead is small relative to useful computation.
Increasing SPR from 4 to 64 at $|\mathcal{A}_\text{raise}|{=}3$ grows the tree to $15.0\times$ its original size (from 57 to 853 nodes), while flop time grows $8.0\times$ (from 6.79 to 54.47~ms), indicating improved parallel efficiency on larger trees due to better load balancing across threads.

\begin{table}[!ht]
\centering
\caption{Per-street CFR pipeline time across raise action count and SPR.}
\label{tab:action_spr}
\begin{tabular}{rrrrrrrr}
\toprule
$|\mathcal{A}_\text{raise}|$ & \textbf{SPR} & \textbf{Nodes} & \textbf{InfoSets} & \textbf{Preflop} & \textbf{Flop} & \textbf{Turn} & \textbf{River} \\
\midrule
2 &  4 &      45 &  105{,}840 & $0.67{\pm}0.04$ & $4.67{\pm}0.60$ & $4.45{\pm}0.59$ & $3.04{\pm}0.10$ \\
2 & 16 &     135 &  317{,}520 & $1.48{\pm}0.03$ & $9.08{\pm}0.30$ & $8.12{\pm}0.36$ & $6.70{\pm}0.18$ \\
2 & 64 &     415 &  976{,}080 & $1.94{\pm}0.13$ & $25.64{\pm}0.07$ & $24.93{\pm}0.07$ & $25.35{\pm}0.33$ \\
\midrule
3 &  4 &      57 &  134{,}064 & $0.72{\pm}0.03$ & $6.79{\pm}0.60$ & $6.40{\pm}0.52$ & $4.36{\pm}0.12$ \\
3 & 16 &     195 &  458{,}640 & $1.79{\pm}0.23$ & $16.09{\pm}0.16$ & $15.29{\pm}0.24$ & $12.84{\pm}0.33$ \\
3 & 64 &     853 & 2{,}006{,}256 & $3.48{\pm}0.22$ & $54.47{\pm}0.40$ & $52.09{\pm}0.25$ & $47.04{\pm}0.25$ \\
\bottomrule
\end{tabular}
\end{table}

\paragraph{Preflop bucket count.}
To understand the computational headroom from abstraction, we evaluate per-iteration time under varying bucket counts $B=\{50,169\}$ 
in Table~\ref{tab:bucket_ablation}. At $B{=}50$, total per-iteration time drops to roughly 53~ms (vs.\ ${\sim}157$~ms at lossless), a $3.0\times$ further reduction on top of multi-threading.

\begin{table}[h]
\centering
\caption{Per-iteration CFR pipeline time (ms) vs.\ bucket count $B$.}
\label{tab:bucket_ablation}
\begin{tabular}{rrrrr}
\toprule
$B$ & \textbf{Preflop} & \textbf{Flop} & \textbf{Turn} & \textbf{River} \\
\midrule
50   & $1.68{\pm}0.18$  & $19.44{\pm}0.34$ & $18.37{\pm}0.26$ & $13.52{\pm}0.11$ \\
169  & $3.48{\pm} 0.22$  & $35.58{\pm}0.24$ & $34.02{\pm}0.17$ & $25.38{\pm}0.21$ \\
\bottomrule
\end{tabular}
\end{table}


\paragraph{Convergence comparison.}
We compare Parallel CFR against PokerRL~\citep{steinberger2019pokerrl}, one of the most efficient open-source vectorized CFR implementations for poker.
For a fair comparison, both solvers operate on the same river subgame tree with $93$ game tree nodes, and we disable the features uniquely supported by Parallel CFR (information abstraction, pruning, and depth-limited solving), so that both solvers perform pure endgame solving on an identical game.

\begin{figure}[h]
\centering
\includegraphics[width=0.7\columnwidth]{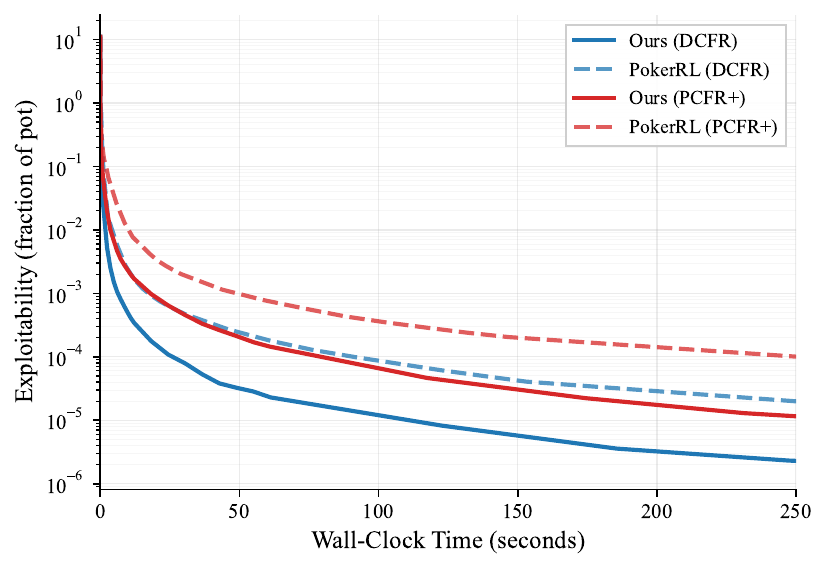}
\caption{Wall-clock convergence comparison between Parallel CFR and PokerRL on a river subgame. Solid lines: Parallel CFR (ours); dashed lines: PokerRL. Both solvers run two CFR variants (DCFR, PCFR+) on the same game tree with no information abstraction or pruning.}
\label{fig:convergence}
\end{figure}

We evaluate two state-of-the-art CFR variants: DCFR~\citep{brown2019solving} and Predictive CFR+~\citep{farina2021faster}.
Figure~\ref{fig:convergence} shows exploitability as a function of wall-clock time for Parallel CFR versus PokerRL. At the same wall-time budget, Parallel CFR reaches $\sim7\times$ lower exploitability than PokerRL across both CFR variants. These results demonstrate that Parallel CFR substantially outperforms existing open-source alternatives even when its advanced features are disabled, confirming both the algorithmic correctness and practical efficiency of our parallel pipeline. Note that this comparison disables pruning and abstraction because PokerRL does not support them; when these techniques are enabled, prior work~\citep{li2025evpa} has shown that they can accelerate CFR convergence by up to two orders of magnitude, further widening the gap.

\section{Conclusion}
\label{sec:conclusion}

We presented Parallel CFR, a parallelization framework for real-time depth-limited CFR solving that decomposes each iteration into seven stages with two orthogonal dimensions of parallelism.
Parallel CFR natively integrates GPU-batched leaf evaluation, online pruning, information abstraction, and advanced CFR variants into a unified heterogeneous CPU--GPU pipeline.
On Heads-Up No-Limit Texas Hold'em, Parallel CFR achieves per-iteration times of $47$--$54$~ms on postflop streets using only 5 CPU threads and 1 GPU on a single desktop-class device (NVIDIA DGX Spark), a $3.3$--$3.4\times$ speedup over the single-threaded baseline. Moreover, a direct convergence comparison against an efficient poker solver confirms that Parallel CFR achieves $\sim 7\times$ lower exploitability at the same wall-time budget. Together, these results enable hundreds of CFR iterations within a typical real-time decision budget of a few seconds, bringing competitive real-time solving from datacenter-scale infrastructure to compact consumer-grade hardware.

\paragraph{Limitations and future work.}
Our parallelization focuses on a single CFR solve instance.
Combining with distributed computing could provide additional scaling. Exploring GPU-native implementations of the full pipeline and asynchronous CPU--GPU overlap are promising directions for further reducing per-iteration latency.

\bibliographystyle{plainnat}
\bibliography{references}

\end{document}